\documentstyle{europhys}

\newif\ifboo \boofalse

\begin{document}
%
%
%
\euro{}{}{}{1999}
\Date{1999}
\shorttitle{A. D. Kent {\it et al.} Low temperature magnetic hysteresis in Mn$_{12}$}
\title{Low temperature magnetic hysteresis in Mn$_{12}$  acetate \\ single crystals }
\author{Andrew D. Kent\inst{1,*}, Yicheng Zhong\inst{2}, Louisa Bokacheva\inst{1}, Daniel Ruiz\inst{3}, 
David N. Hendrickson\inst{3}, and M. P. Sarachik\inst{2}}
\institute{
     \inst{1} Department of Physics, New York University, 4 Washington Place\\
New York, NY 10003 USA\\
     \inst{2} Physics Department, City College of the City University of New
York\\ New York, NY 10031 USA\\
     \inst{3} Department of Chemistry and Biochemistry, University of California at San 
Diego\\ La Jolla, CA 92093 USA}
\rec{7 July 1999}{}
\pacs{
\Pacs{75}{45$+$j}{Macroscopic quantum phenomena in magnetic systems}
\Pacs{75}{60 Ej }{Magnetisation curves, hysteresis, Barkhausen and related effects}
\Pacs{75}{50 Tt}{Fine-particle systems}      }
\maketitle
\begin{abstract}
Precise magnetic hysteresis measurements of small single crystals of Mn$_{12}$ acetate of 
spin 10 have been conducted down to $0.4$ K using a high sensitivity Hall magnetometer. 
At higher temperature ($> 1.6$ K) step-like changes in magnetization are observed at 
regularly spaced magnetic field intervals, as previously reported. However, on lowering 
the temperature the steps in magnetization shift to higher magnetic fields, initially 
gradually. These results are consistent with the presence of a second order uniaxial 
magnetic anisotropy, first observed by EPR spectroscopy, and thermally assisted 
tunnelling with tunnelling relaxation occurring from levels of progressively lower energy 
as the temperature is reduced. At lower temperature an abrupt shift in step positions is 
found. We suggest that this shift may be the first evidence of an abrupt, or first-order, 
transition between thermally assisted and pure quantum tunnelling, suggested by recent 
theory.
\end{abstract}
%
%
%
%
%
\section{Introduction}
The high spin ($S=10$) molecular magnets Mn$_{12}$ acetate and Fe$_8$ have become 
prototypes for the study of the transition from classical superparamagnetism to quantum 
tunnelling of mesoscopic spins. Much of the recent interest in these materials has been 
stimulated by the observation of a remarkably regular series of steps and plateaus in the 
magnetic hysteresis loops of Mn$_{12}$ at low temperature (below a blocking 
temperature of $3$ K), first in oriented powders \cite{1} and shortly thereafter in single 
crystals \cite{2}. These results indicate that the relaxation rate of the magnetization 
toward equilibrium is greatly enhanced at well-defined intervals of magnetic field. These 
observations have been interpreted within a simple effective spin Hamiltonian for these 
molecules and a model of thermally assisted tunnelling of the magnetization, first 
suggested in reference \cite{3}. This model describes a regime intermediate between 
thermal activation over the anisotropy barrier (superparamagnetism) and pure quantum 
tunnelling ($T=0$) in which both thermal activation and quantum tunnelling are 
important to the magnetization reversal. 

Mn$_{12}$ has subsequently been studied extensively and by a variety of techniques. 
Notably, both EPR \cite{4} and inelastic neutron spectroscopy \cite{5,6,7} have been used 
to independently determine the parameters for the spin Hamiltonian of Mn$_{12}$. These 
have shown that higher order terms in the Hamiltonian--not considered in the analysis 
thus far of magnetic hysteresis data--are necessary to fit the spectra. Surprisingly few 
experiments have been conducted at lower temperature in Mn$_{12}$  to study the 
transition to pure quantum tunnelling behaviour, in contrast to important studies of this 
type in Fe$_8$  \cite{8,9}. Some earlier experiments showed the appearance of Òmagnetic 
avalanchesÓ at lower temperature, that is, rapid and uncontrolled magnetization 
switching to its saturation value, which precluded controlled low temperature relaxation 
studies \cite{10}. Later studies using cantilever magnetometry revealed several new 
higher field magnetization steps at lower temperature, consistent with the model of 
thermally assisted tunnelling \cite{11}.

In this letter we present precise low temperature high field magnetic hysteresis 
measurements on Mn$_{12}$ which reveal two important new aspects of the 
magnetization reversal. First, on lowering the temperature below $1.6$ K, we find that 
steps in magnetization shift gradually to higher magnetic field, consistent with the 
presence of a second order (fourth power) uniaxial magnetic anisotropy constant determined by EPR 
spectroscopy \cite{4}  and, recently, by precise inelastic neutron scattering measurements 
\cite{5,6}. Second, at lower temperature an intriguing abrupt shift in step position is 
found. We suggest that this shift may be due to an abrupt transition between thermally 
assisted and pure quantum tunnelling in Mn$_{12}$ acetate, first predicted theoretically by 
Chudnovsky and Garanin \cite{12}. We also discuss other possible origins of this 
observation.

\section{Background}
Mn$_{12}$ acetate crystals have a tetragonal lattice ($a=1.73$ nm and $b=1.24$ nm) of 
molecules with $12$ interacting mixed valent Mn ions with a net ground state spin of 
$10$, $S=10$ \cite{13}. Thus each molecule has $2S+1=21$ magnetic levels, labeled by 
quantum number $m$ ($m=-10,-9,-8,...,10$). The molecules have a strong uniaxial 
magnetic anisotropy energy, and to a good approximation the spin Hamiltonian can be 
written:
\begin{equation}
H= -DS_z^2-BS_z^4 - g_z\mu_BH_zS_z + H'
\label{eqn1}
\end{equation}
The parameters D and B have been determined first by EPR spectroscopy \cite{4} and 
now very accurately by inelastic neutron spectroscopy \cite{6} to be $D = 0.548(3)$ K, 
B=$1.173(4) \times 10^{-3}$ K, and $g_z$ is estimated to be $1.94(1)$ \cite{4}. Spin 
alignment is favored to be up ($m=10$) or down ($m=-10$) along the z-axis. The energy 
barrier between up and down states is approximately $67$ K. The third term is the 
Zeeman energy for fields applied parallel to the easy axis. $H'$ represents small terms 
which break the axial symmetry and, hence, produce tunnelling. These are due to 
transverse fields (i.e., terms like $H_xS_x$), and higher order magnetic anisotropies, the 
lowest order form allowed by tetragonal symmetry being ($S_+^4 + S_-^4$). By itself, this 
last term would lead to a tunnelling selection rule with $\Delta m=4i$, with $i$ an integer. 
Since this is not found experimentally, it is likely that transverse fields due to hyperfine, 
dipolar fields ($\sim 0.1$ K) and/or an external applied field (such as due to a small 
misalignment between the applied field and the z-axis) are most important to mixing the 
m-levels and producing tunnelling.

The steps observed in magnetic hysteresis measurements, their temperature dependence 
and magnetization relaxation experiments provide strong evidence for thermally assisted 
tunnelling. Within this model magnetization reversal occurs by tunnelling from thermally 
excited magnetic sublevels (i.e, $m=9,8,7,...,-8,-9$) at magnetic fields at which these levels 
are in resonance with levels on the opposite side of the anisotropy barrier (inset Fig. 1). 
From (1) levels $m$ and $m'$ have the same energy when:
\begin{equation}
H=nH_o\left[1 + \frac{2B}{D}\left(\left(m-\frac{n}{2}\right)^2 +
\frac{n^2}{4}\right)\right]
\label{equation}
\end{equation}
where $n=m+m'$ is the step index, $H_o=D/g_z\mu_B=0.42$ T is a field ``quantum'' and 
$m$ is the escape level from the metastable well. At these magnetic fields, the 
magnetization relaxation can occur on measurement time scales and give rise to the
step-like changes in magnetization. Otherwise, the relaxation rate is slower, leading to plateaus 
in the magnetic hysteresis loop. The relaxation rate from any level is proportional to the 
product of the thermal occupation probability of that level and the probability for 
quantum tunnelling from the level, and thus should increase exponentially with 
temperature, as found in experiments above $2$~K \cite{13,14}. Since the tunnelling 
probability is intrinsically small for lower lying levels (with large $\Delta m=m-m'$), 
larger magnetic fields are necessary at lower temperature to produce observable 
tunnelling relaxation, also, as seen in experiments. 

\begin{figure}[t]
\centerline{\epsfxsize=8 cm \epsfbox{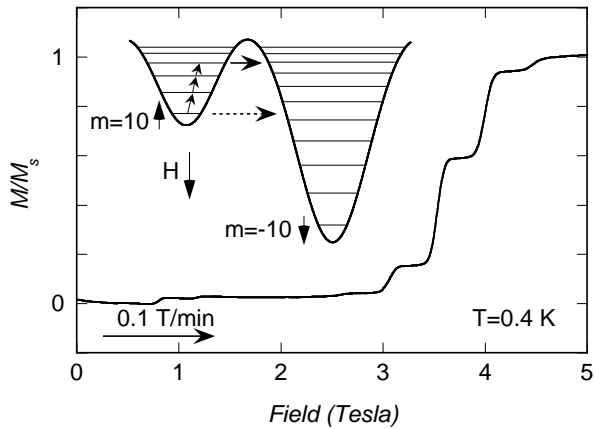}}
\caption{ Magnetic hysteresis measured on a small Mn$_{12}$ single crystal with a micro-Hall
magnetometer at $0.4$ K. The field is applied along the easy magnetic axis (within a 
few degrees) and the field ramp rate is $0.1$ T/min. The inset shows a schematic of the 
magnetic potential in an applied field, with the horizontal axis denoting the angle between 
the spin direction and the easy axis. Escape may occur via pure quantum tunnelling 
(dashed horizontal arrow) or thermally assisted tunnelling (vertical arrows up the 
metastable well) and solid arrow across the energy barrier. The effect of the second order 
uniaxial magnetic anisotropy is also illustrated. Levels on opposite sides of the barrier 
cross at different magnetic field.}
\label{fig1}
\end{figure}

Importantly, within this model, tunnelling relaxation occurs from small group of 
quasilevels in the metastable well--the escape levels, $m_{esc}$. This is because the 
tunnelling probability increases exponentially with energy $E$, as the effective barrier 
height becomes lower, while the thermal occupation probability decreases exponentially 
with energy, $exp(-E/kT)$. Note also from equation (2) that for a given step index n the 
fields at which steps occur depend on the escape levels. Larger fields are necessary to 
bring lower lying levels, i.e., larger $m$ levels, into resonance (as generally, $m>n/2$), so 
that a shift in step position to higher fields signals tunnelling from states deeper in the 
metastable potential well (i.e., larger $m_{esc}$). Finally, dipolar interactions between 
clusters, interactions with nuclear spins \cite{15} and spin-phonon interactions are 
essential to a quantitative and microscopic understanding of the relaxation \cite{16,17} 
such as, the observation of non-exponential relaxation \cite{8} and both the linewidth and 
form of the relaxation peaks \cite{18}.

\section{Experiment}
The magnetization of small single crystals of Mn$_{12}$-acetate in the form of a 
parallelepiped ($50 \times 50 \times 300 \mu m^3$) was measured using a high 
sensitivity micro-Hall effect magnetometer \cite{19}. Like a micro-superconducting 
quantum interference device ($\mu$-SQUID) \cite{20}, this magnetometer measures a 
magnetic field induced by the crystal's magnetization. The measurements are done in a rf 
shielded automated high-field Helium $3$ system, in which careful attention has been 
paid to reducing electrical noise and to thermalizing the sample. Temperature is measured 
both at the cold stage of the cryostat and with a small resistance thermometer mounted 
within a few mm of the sample. These measurements are always within $50$ mK of one 
another. Fig. 1 shows a typical portion of a hysteresis curve measured at $0.4$~K starting 
from a demagnetized state, $M=0$, and measured at a ramp rate of $0.1$ T/min, with the 
field along the easy axis (within a few degrees). Prominent step-like changes in 
magnetization are observed at fields between $3$ and $5$ Tesla.

\begin{figure}[t]
\centerline{\epsfxsize=8 cm \epsfbox{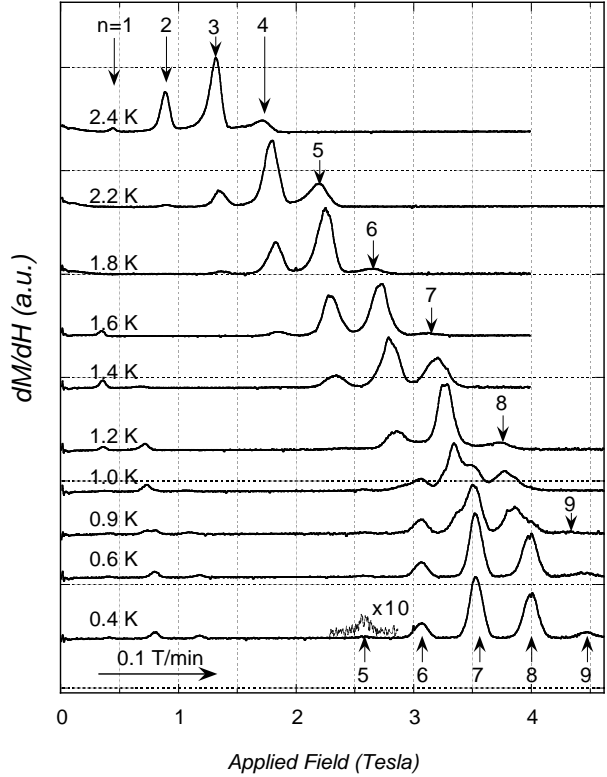}}
\caption{Derivative of the magnetization with respect to applied field versus field at 
temperatures from $2.4$ K to $0.4$ K. Data are acquired from $M=0$ at constant field 
ramp rate of $0.1$ T/min. The small peaks at low temperature and fields ($H/H_o <3$) 
are due to a small magnetic impurity phase in the crystal, discussed in reference 
\cite{24}. Data near the $n=5$ peak at $0.4$ K is multiplied by $10$ and offset so that it 
is visible on the plot.}
\label{fig2}
\end{figure}

Fig. 2 shows the derivative of the magnetization curves $dM/dH$ versus applied field at 
different temperatures. Peaks in $dM/dH$ correspond to maxima in the magnetization 
relaxation rate at that applied field, sample magnetization and measurement temperature. 
Examining the data from high to low temperature; first, above $1.6$ K the data are in 
good accord with previous experiments \cite{1,2}. Peaks appear at approximately equally 
spaced field intervals ($\sim 0.45$ T) and their amplitude is a strong function of the 
temperature. As the temperature is reduced, higher numbered maxima in $dM/dH$ 
appear, while lower field peaks decrease in amplitude, again, consistent with the model of 
thermally assisted tunnelling. Second, on lowering the temperature peaks shift 
continuously to higher fields. For instance, peak $n=5$, at $2.2$ K is at $2.20$ T and by 
$1.4$ K has shifted to $2.33$ T. Third, and most intriguing, at lower temperature ($T<1.2$ 
K) peaks in $dM/dH$ shift dramatically in position as a function of temperature. This is 
well illustrated by the behaviour of the $n=7$ peak as the temperature is reduced. This 
peak first appears at $1.6$ K at $H=3.10$ T, grows in amplitude and shifts to significantly 
higher fields on lowering the temperature and, at $1.0$ K, abruptly develops a high field 
shoulder. On slightly lowering the temperature to $0.9$ K, ``spectral'' weight is transferred 
into this shoulder and at the lowest temperature the peak remains fixed in position. This 
peak has shifted to $3.53$ T, by a full field quantum $H_o$, in this temperature interval. 
Shifts in peak position of this order are seen for all the steps observed at low temperature 
($5\leq n \leq 9$). Finally, note that at $0.6$ K and lower temperature, the maxima 
remain fixed in field and approximately constant in amplitude. 

The dependence of the $dM/dH$ peak positions on temperature are summarized in Fig. 
3. Here the peak positions, in internal field ($H_{int}=H+4 \pi M$) divided by the field 
quantum, $H_o$, are plotted versus temperature. Note that peaks initially shift gradually 
to higher magnetic fields as the temperature is lowered. Between $0.6$ K and $1.2$~K the 
peak positions shift abruptly, with higher step indices changing position at lower 
temperature. The solid vertical line demarcates the approximate temperature at which 
these sudden shifts in step position occur. Below this line the step positions are 
independent of temperature.

\begin{figure}[t]
\centerline{\epsfxsize=8  cm \epsfbox{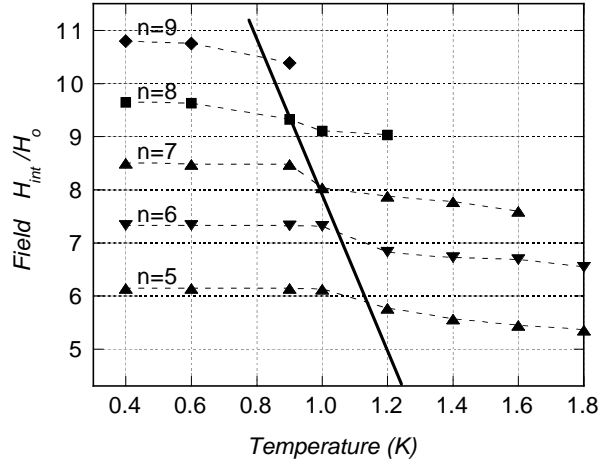}}
\caption{Relaxation peak positions, $H_{int} =H+4 \pi M$ \cite{25}, normalized to 
$H_o$, the field quantum,  versus temperature. The solid line shows the approximate 
temperature below which peak positions are temperature independent. }
\label{fig3}
\end{figure}

\section{Discussion}
These experiments are the first to show that the steps in magnetization of Mn$_{12}$ 
are not always at regular magnetic field intervals and that steps shift to higher magnetic fields as 
the temperature is lowered. The data are consistent with the presence of a second order 
(fourth power) uniaxial term in the magnetic anisotropy of Mn$_{12}$ (eqn. 1) and indicate that lower lying 
magnetic sublevels dominate the tunnelling relaxation as the temperature is reduced (eqn. 
2). 

We can clearly distinguish the physical behaviour in two different temperature regimes. 
At higher temperature (above the solid line in Fig. 3) the step positions shift gradually 
with temperature. This is the regime of thermally assisted tunnelling, where the magnetic 
escape is from thermally excited magnetic levels. We associate shifts in step positions with 
incremental changes in these levels, such as from $m$ to $m+1$, as the temperature is 
reduced and/or changes in the relative importance of a few levels, which contribute ``in 
parallel'' to the magnetization relaxation at a given temperature. Fig. 4 shows the energy 
levels versus field for the Mn$_{12}$ spin Hamiltonian (eqn. 1). The vertical solid lines  
indicate the level coincidences important to the magnetic relaxation in this temperature 
and field range. For example, for 
$n=6$, the step positions we find are consistent with transitions from $m=8$ to $m'=-2$ 
and $m=7$, to $m'=-1$.

The second regime is at low temperature (below the solid line in Fig. 3) in which the 
position of peaks in $dM/dH$ are independent of temperature. This is consistent with 
magnetization relaxation being of a pure quantum nature, i.e. tunnelling escape occurring 
from the lowest level in the metastable well, $m=10$. At $0.6$ K and below, the amplitude 
of the peaks are also temperature independent this is additional evidence for a quantum 
regime in Mn$_{12}$, as it indicates that the relaxation of the magnetization in our 
measurement time window has become temperature independent.

\begin{figure}[t]
\centerline{\epsfxsize=10 cm \epsfbox{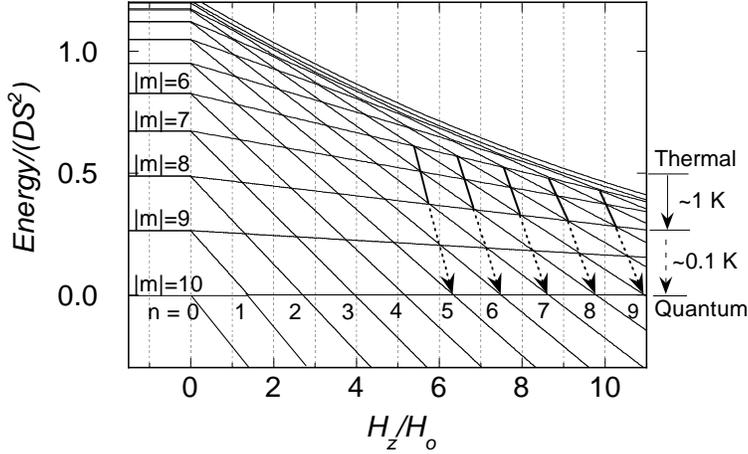}}
\caption{ Energy level diagram versus field for Mn$_{12}$-acetate obtained by
diagonalizing the 
Hamiltonian [eqn. 1]. Energy is measured relative to the lowest
lying state in the metastable well. The levels important to the low temperature relaxation are 
indicated. The abrupt shifts in the magnetization steps with decreasing 
temperature are consistent with the escape levels changing as illustrated by the dashed 
arrows from $m=8$ to $m=10$.}
\label{fig4}
\end{figure}

The most striking feature of these data is the abrupt shift in step position, observed at the 
boundary between these temperature regimes. This shift suggests that different levels 
become important to the tunnelling relaxation in a narrow temperature interval. 
The shift in peak positions of Fig. 
3 is consistent with the change in levels responsible for tunnelling illustrated by the 
dashed arrows in Fig. 4--$m_{esc}$ changes by $2$ in an interval of
$0.1$ to $0.2$~K. The abrupt nature of this transition is evident directly from the 
magnetic hysteresis data in Fig. 2. For example, that the shoulder which develops for the n=7 
peak at 1.0 K (Fig. 2) indicates that metastable levels $m=8$ ($m'=-1$) and $m=10$ ($m'=-3$), both 
contribute to the magnetic relaxation at this temperature but at different easy axis 
magnetic fields. 

We now speculate as to the origin of the abrupt in shift step position with temperature. 
The most interesting possibility is that the abrupt shift in peak position we observe is 
evidence for a first-order transition between thermally assisted and pure quantum 
tunnelling, as suggested in ref. \cite{12}. In this theory it is shown that for a small uniaxial 
magnetic particle the energy of the quasilevels in the metastable magnetic well which 
dominate the magnetic escape need not be a smooth function of temperature. Larkin and 
Ovchinnikov called the smooth transition from classical thermal activation to pure 
quantum tunnelling a second-order transition \cite{21}, regarding the energy of escape as 
analogous to an order parameter in a phase transition problem. For small transverse 
fields, Chudnovsky and Garanin find that the transition can be first-order with certain 
energy levels in the metastable well being skipped entirely as the temperature is varied. 
They considered both a large uniaxial spin in a quasiclassical approximation \cite{12} as 
well as small spins ($S \sim 10-100$) with a discrete level spectrum \cite{22,23}, as in 
Mn$_{12}$.

There may, of course, be other explanations for the observed shift in relaxation peaks 
with temperature. A tacit assumption we make is that peaks in $dM/dH$ correspond to 
maxima in the relaxation rate at a given field (including the internal field). Then the 
maximum shift in relaxation rate maxima due to the internal fields is 
about $4 \pi M \simeq0.1$ 
T, which is smaller than the changes that we observe ($ \approx 0.4$ T). However, it may 
not be possible to account for the internal fields in this average way. For instance, the 
distribution of internal fields throughout the crystal likely changes in a complex manner 
during our field sweep experiments. 

It is also possible that sample heating plays a role. For example, if the sample were not in 
thermal equilibrium with the thermometers during the measurements and actually at a 
higher temperature, this would explain the temperature independent behaviour observed 
below $0.6$ K. Sample heating may also play another role. Relaxation of the 
magnetization leads to strong dissipation, which leads to sample heating which, in turn, 
leads to enhanced magnetization relaxation. This positive feedback is at the origin of the 
magnetic avalanches reported in ref. \cite{10}. Perhaps, this positive feedback could 
produce the shoulder-like structures we observe on certain $dM/dH$ peaks in Fig. 2 
($0.9$ and $1.0$ K). We estimate that the maximum heat generated in these experiments 
is $H(dM/dt) =H(dM/dH)(dH/dt)=1 $ nW, which is of the same order as the heat 
dissipated in our Hall magnetometer ($2$ nW).  Nonetheless, we have sometimes 
observed magnetic avalanches at higher sweep rates ($0.4$ T/min)Ñso we cannot 
completely rule out this possibility. Finally, while the peak shifts we observe are abrupt, 
the transition could still be continuous but occur over a narrow temperature interval. 
Further detailed experimentation and modeling are likely to clarify this situation.

In summary, we have presented new data which suggest the transition between 
thermally assisted and pure quantum tunneling in Mn$_{12}$ may be abrupt, or first 
order. Importantly, these results show that magnetization relaxation and magnetic 
hysteresis measurements may be used to do a new type of spectroscopy of the levels 
important to magnetic escape in Mn$_{12}$. Further experiments and modeling will 
undoubtedly lead to a better fundamental understanding of this transition between 
thermally assisted and pure quantum tunneling.

\section{Acknowledgements}
This work was supported by at NYU by NSF-INT grant 9513143 and NYU, at CCNY by 
NSF grant DMR-9704309 and UCSD by NSF grant DMR-9729339.
\\
\\
$^*$Corresponding author: andy.kent@nyu.edu


\begin{thebibliography}{99}
\bibitem{25} The maximum correction is $4 \pi M_s=0.063$ T and is relatively small on 
the scale at which the data is plotted in Fig. 2.
\end{thebibliography}
\end{document}